\UseRawInputEncoding

\documentclass[11pt, oneside]{article}   	

\pdfoutput=1
\usepackage{jheppub}

\usepackage[english]{babel}
\usepackage[T1]{fontenc}
\usepackage{lmodern}

\usepackage{graphicx}	

\usepackage{amsmath}								
\usepackage{color}
\usepackage{mathtools}
\usepackage{amssymb}

\usepackage{hyperref}

\hypersetup{
    bookmarks=true,%
    colorlinks,%
    citecolor=blue,%
    filecolor=black,%
    linkcolor=blue,%
    urlcolor=blue
}

\usepackage{overpic}
\usepackage{array}
\usepackage{rotating}

\usepackage{amssymb}

\let\emptyset\varnothing

\title{Pure de Sitter space and the island moving back in time}

                                           \author{Watse Sybesma}
                                           \affiliation{Science Institute
                                           University of Iceland \\Dunhaga 3, 107 Reykjav\'{i}k, Iceland.}

                                           \emailAdd{watse@hi.is}

\abstract{
Observers in de Sitter space can only access the space up to their cosmological horizon. 
Assuming thermal equilibrium, we use the quantum Ryu-Takayanagi or island formula to compute the entanglement entropy between the states inside the cosmological horizon and states outside, as a function of time. We obtain a Page curve that is bound at a value corresponding to the Gibbons-Hawking entropy.  

At this transition an `island' forms, which is in a significantly different location as compared to when considering black hole horizons and even moves back in time. 
These differences turn out to be essential for non-violation of the no-cloning theorem in combination with entanglement wedge reconstruction. 
This consideration furthermore introduces the need for a scrambling time, the entropy dependence of which turns out to coincide with what is expected for black holes.

The model we employ has classically pure three-dimensional de Sitter space as a solution. We dimensionally reduce to two dimensions in order to take into account semi-classical effects. Nevertheless, we expect the aforementioned qualitative features of the island to persist in higher dimensions.
}

\begin{document}
\maketitle

\section{Introduction}
De Sitter space is currently arguably the most elusive of maximal symmetric spaces. For example, compared to flat space and anti-de Sitter, a solid holographic interpretation or string theoretical realization are still not as well understood, see e.g. \cite{Strominger:2001pn,Witten:2001kn,Maldacena:2002vr,Pilch:1984aw,Danielsson:2018ztv,Obied:2018sgi,Gautason:2018gln} and references therein.
This poses a challenge, since we have strong indications to believe that our current universe at large scales can be approximated by de Sitter space and furthermore, also early universe inflation can be approximated by this space, see e.g. \cite{Baumann:2014nda} and references therein.

An enigmatic feature of de Sitter space is the appearance of causally disconnected regions. 
Each observer can only access parts of the universe, bounded by their cosmological horizon. At the semi-classical level this horizon, much like in the case of a black hole, emits and reabsorbs radiation, called Gibbons-Hawking radiation \cite{Gibbons:1977mu}. 
The similarity does not end here, as one can associate entropy to the cosmological horizon through the Bekenstein-Hawking formula, called Gibbons-Hawking entropy \cite{Gibbons:1977mu}.
However, a key difference is that, in contrast to black holes for which we believe that this entropy counts states, the cosmological horizon is an observer dependent property and its entropy is thus believed to measure the observer's ignorance about what lies beyond the cosmological horizon. 
See reviews \cite{Spradlin:2001pw,Anninos:2012qw,Bousso:2002fq} for more details.
\begin{figure}[h!!]
	\begin{center}
	\vspace{0.25cm}	
		\begin{overpic}[width=0.5\textwidth,]{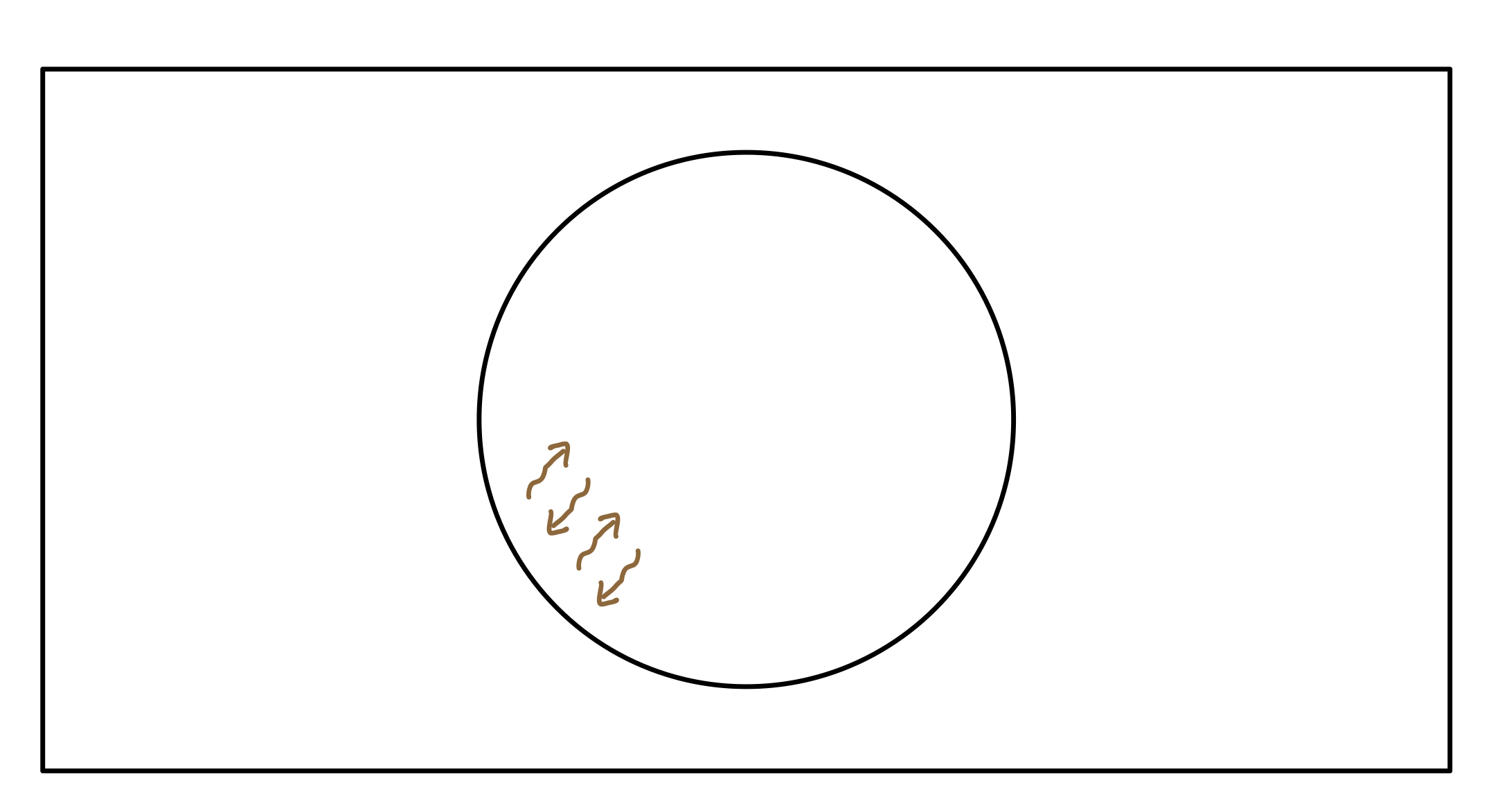}
		
			\put (38,52) {\footnotesize{de Sitter space}}
			\put (43,19) {$\leftarrow$\footnotesize{radiation}}
			\put (45,30) {\footnotesize{inside}}
			\put (44,26) {\footnotesize{horizon}}			
			\put (10,30) {\footnotesize{outside}}
			\put (10,26) {\footnotesize{horizon}}
			\put (70,25) {\footnotesize{$\leftarrow$}}	
			\put (74,27) {\footnotesize{cosmological}}	
			\put (79,23.5) {\footnotesize{horizon}}			
		\end{overpic}
	\end{center}
	\vspace{-0.5cm}
\caption{
The goal of this paper is to study the entanglement entropy between the states inside the cosmological horizon and the states outside. Using the island formula we find the Page curve given in Figure \ref{fig:pagecurve}.
}
\label{fig:goal} 
\end{figure}
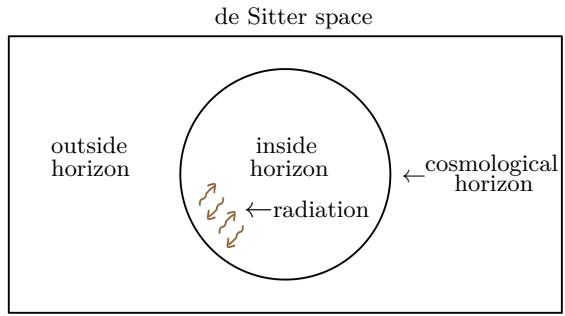

In this paper we point out some more similarities between black holes and pure de Sitter space.\footnote{
With pure we refer to the fact that at classical level there is only a cosmological constant and no matter.}
We show that the entanglement entropy between states within the finite region bounded by an observer's cosmological horizon and the states beyond the cosmological horizon saturates at a value given by the Gibbons-Hawking entropy, see Figure \ref{fig:goal}. 
We furthermore show that one can associate a scrambling time in the sense of the Hayden-Preskill protocol \cite{Hayden:2007cs} to the cosmological horizon, which seems to be in line with e.g. \cite{Susskind:2011ap,Aalsma:2020aib,Nomura:2011dt,Geng:2020kxh}.

To arrive at these results we have employed the quantum Ryu-Takayanagi or island formula \cite{Ryu:2006bv,Hubeny:2007xt,Faulkner:2013ana,Barrella:2013wja,Engelhardt:2014gca,Penington:2019npb,Almheiri:2019psf,Almheiri:2019hni}. The island formula has been used to reproduce the Page curve \cite{Page:1993wv,Page:2013dx} for various black holes \cite{Penington:2019npb,Almheiri:2019psf,Almheiri:2019hni,Almheiri:2019yqk,Almheiri:2019psy,Gautason:2020tmk,Anegawa:2020ezn,Hashimoto:2020cas,Hartman:2020swn,Hollowood:2020cou,Dong:2020uxp,Chen:2020tes,Hartman:2020khs,Balasubramanian:2020hfs,Balasubramanian:2020xqf,Balasubramanian:2020coy,Alishahiha:2020qza,Chen:2020jvn,Geng:2020qvw}, which in essence implies that at late time entanglement entropy in those systems stops growing and is in fact bounded by an amount of entropy matching the Gibbons-Hawking entropy. This transition is directly related to the appearance of an island region behind the horizon. Although the island formula was initially motivated by holography, one can also reach the same conclusion using the replica trick and its extensions, see e.g. \cite{Almheiri:2019qdq,Penington:2019kki}. Islands in cosmology have been studied in \cite{Dong:2020uxp,Chen:2020tes,Hartman:2020khs,Balasubramanian:2020xqf,Balasubramanian:2020coy}. Entanglement entropy in de Sitter space has been studied in Refs. \cite{Sato:2015tta,Nomura:2017fyh,Narayan:2017xca,Arias:2019pzy,Narayan:2020nsc} from a holographic point of view. 

\begin{figure}[h!!]
	\begin{center}
	\vspace{0.25cm}	
		\begin{overpic}[width=0.8\textwidth]{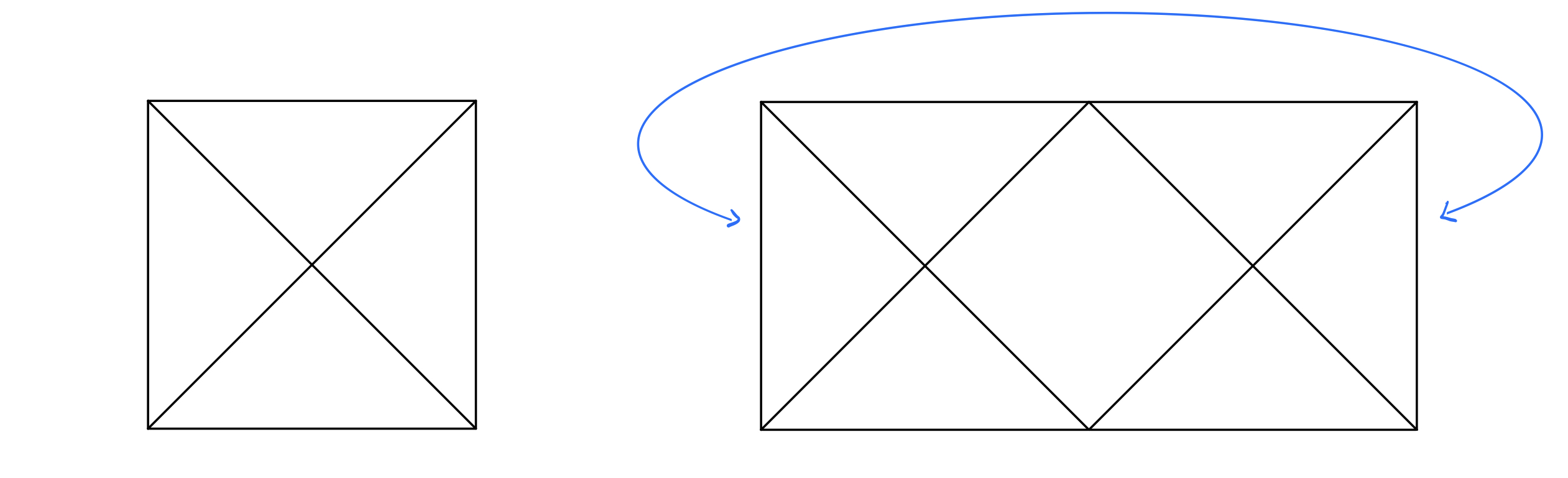}
		
			\put (31.5,20) {\rotatebox{-90}{\footnotesize{North pole}}}
			\put (6,7.5) {\rotatebox{90}{\footnotesize{South pole}}}
			\put (65,27) {\footnotesize{identify}}
		\end{overpic}
	\end{center}
	\vspace{-0.5cm}
\caption{
Left: Causal diagram of pure de Sitter space beyond two dimensions (more details in Section \ref{sec:model}). The vertical lines are identified as the North and South pole. Right: Causal diagram of pure de Sitter space in two dimensions. Horizontal slices represent a circle. In both figures the diagonal lines indicate cosmological horizons and horizontal lines $\mathcal{I}^{\pm}$.
}
\label{fig:penroses} 
\end{figure}

To fully utilize the island formula one has to provide a gravitational model with semi-classical corrections. To comply with this requirement, we study pure de Sitter space in three dimensions by doing a dimensional reduction to two dimensions such that we are able to take into account semi-classical corrections using the conformal anomaly along the lines of Refs. \cite{Christensen:1977jc,Callan:1992rs}. We would like to emphasize that the resulting two-dimensional causal diagram is different from the pure two-dimensional de Sitter space causal diagram, see Figure \ref{fig:penroses}, and is in fact an avatar for three-dimensional pure de Sitter space. Technically this can be brought back to the fact that in the case under consideration in this paper a certain topological term is absent which in the other case encodes the entropy of a higher dimensional extremal limit. We will make this concrete in the main body of the text.

The island we find in pure de Sitter space exhibits two features which, to our knowledge, contrast the behavior of islands appearing in the context of black holes. Firstly, the island does not arise near the future horizon. Secondly, the island moves back in time.\footnote{To be more precise, the island moves in the same direction as the usual timelike Killing vector in the South pole, which is downwards. We however choose to invert time there. The important point is that in the current case the island moves downwards in the Penrose diagram, whether for black holes it is found to move upwards.} See Figure \ref{fig:flat1} for a comparison. We will argue that both these features are essential because otherwise the no-cloning theorem would be violated when entanglement wedge reconstruction is applied \cite{Czech:2012bh,Headrick:2014cta,Wall:2012uf,Jafferis:2015del,PhysRevLett.117.021601,Cotler:2017erl,Penington:2019npb}. As these arguments turn out to not be tied to specific dimensions, we speculate that qualitatively this island behavior holds beyond three dimensions.

Apart from the absence of derivations through holography or replica wormholes, the application of the island formula in this paper is speculative as there is no region where gravity is absent. We argue however, that in analogy to flat space where near $\mathcal{I}^{+}$ gravity is weak see e.g. \cite{Gautason:2020tmk}, the gravitational strength within the static patch can be made weak enough to warrant usage of the island formula.

This paper is organized in the following fashion. We start by motivating our way to use the island formula in Section \ref{sec:motivation}. Afterwards we introduce the model in Section \ref{sec:model}. In Section \ref{sec:pagecurve} we evaluate the island formula and present the Page curve. We discuss the importance of the location of the island and the fact that it moves back in time in Section \ref{sec:discussion}. We end with an outlook in Section \ref{sec:outlook}. Throughout this paper we put the speed of light $c$, Boltzmann constant $k_{B}$ and Planck's constant $\hbar$ to unity.
\\
\\
\textbf{Note added:} 
While this work was nearing completion, complementary work discussing -- amongst other topics -- islands in de Sitter space appeared \cite{Chen:2020tes,Hartman:2020khs}. In this same period another work \cite{Balasubramanian:2020xqf} on islands in de Sitter space appeared, which reaches the same conclusion about entanglement entropy of pure de Sitter space capping off, but using a different implementation of the island formula.
\section{Motivation island formula usage and setup}\label{sec:motivation}
In the seminal works on the island formula \cite{Penington:2019npb,Almheiri:2019psf} an evaporating black hole in Anti-de Sitter space is studied. The evaporation occurs due to the fact that radiation is allowed to escape into a coupled heat bath. Using the island formula, the entanglement between the radiation in the bath and the complement, i.e. the gravitational system, is studied. At late times the island formula predicts that the entanglement entropy cannot keep on increasing due to the formation of a so-called island region behind the horizon which is part of the entanglement wedge of the heat bath. This reproduces the Page curve. The validity of this computation can be motivated using the AdS/CFT duality \cite{Maldacena:1997re,Gubser:2002tv,Witten:1998qj} and the replica trick \cite{Almheiri:2019qdq,Penington:2019kki}. 
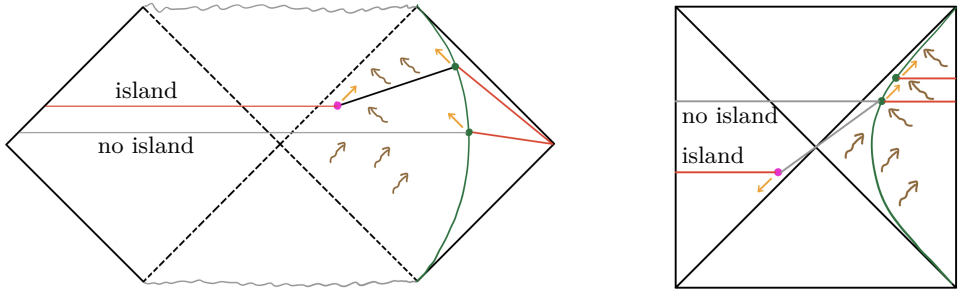
\begin{figure}[h!!]
	\begin{center}
		\begin{overpic}[width=0.85\textwidth]{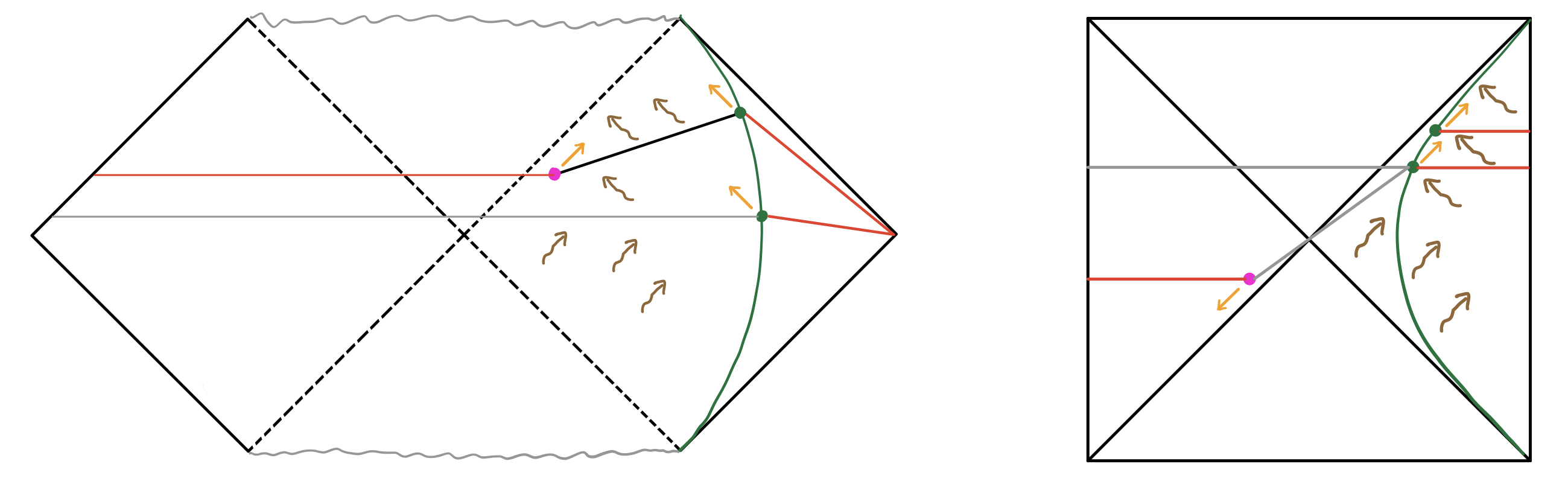}
			\put (13,21.5) {\footnotesize{island}}
			\put (11.25,16) {\footnotesize{no island}}
			\put (70,19) {\footnotesize{no island}}
			\put (70,15) {\footnotesize{island}}
		\end{overpic}
		
	\end{center}
	\vspace{-0.25cm}
\caption{
In both figures the pink dot indicates the location of the extremal surface, the green dot indicates the location of the anchor point on the anchor curve (green line). The orange arrows denote movement directions in time and the brown wavy arrows represent semi-classical radiation. Left: A Penrose diagram of an eternal black hole in flat space is shown \cite{Gautason:2020tmk} where the horizontal gray wavy line denote the location of the singularities and the diagonal dashed lines represent the horizon. Right: Penrose diagram of pure de Sitter beyond two dimensions. The vertical lines are the locations of the poles and can be reached in finite time, the diagonal lines represent the cosmological horizon and the horizontal lines represent $\mathcal{I}^{\pm}$. For more details see Section \ref{sec:model}.
}
\label{fig:flat1} 
\end{figure}

In Ref. \cite{Gautason:2020tmk} an evaporating and eternal black hole in flat space are studied by introducing an anchor curve. The anchor curve can be thought of as a divider of regions, as we will argue. The anchor curve is placed at some large radial location -- far away from the black hole -- and as a result there is effectively flat space on the right hand side and the gravitational system is contained on the left hand side, see Figure \ref{fig:flat1}. In this case the island formula computes the entanglement of the radiation passing through the right hand side of the anchor curve with the complement on the left hand side of the anchor curve. At late time this also yields an island and reproduces the expected Page curve. The holographic motivation of this computation is less obvious than the aforementioned setup \cite{Gautason:2020tmk,Hartman:2020swn}, although the employed model has a known supergravity pedigree, see \cite{Callan:1992rs}.

In the current work we consider pure de Sitter space in three dimensions. Each observer in de Sitter space experiences radiation coming from and being reabsorbed by their cosmological horizon. In analogy to the aforementioned black holes in flat space, we employ an anchor curve. This time the anchor curve will be `hugging' the cosmological horizon at late times, see Figure \ref{fig:flat1}. As such the anchor line divides the space into the interior and the exterior of the cosmological horizon. We are then interested in computing the entanglement between the states on the right hand side of the anchor curve and the complement on the left hand side of the anchor curve. The main theme of this paper is to analyze what occurs in this setup.

This approach has a more conjectural nature than the aforementioned investigations, e.g. due to the fact that in those cases there was always a non-gravitational bath present. However, we might get around this requirement as we will not `store' radiation in the current setup and since we will show that within the static patch gravity can always be made arbitrarily weak.

\section{The model under consideration}\label{sec:model}
We consider a model with a pure three-dimensional de Sitter space as a solution and then we reduce it to two dimensions because this simplifies the semi-classical analysis. This is in the spirit of e.g. Refs. \cite{Achucarro:1993fd,Maxfield:2020ale,Cotler:2019nbi,Maldacena:2019cbz}.
\subsection{Three-dimensional de Sitter space and its two-dimensional reduction}
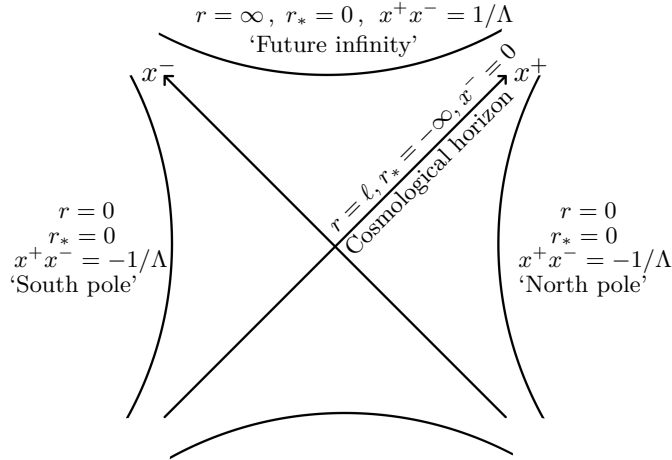
\begin{figure}[h!!]
	\begin{center}
		\begin{overpic}[width=0.4\textwidth]{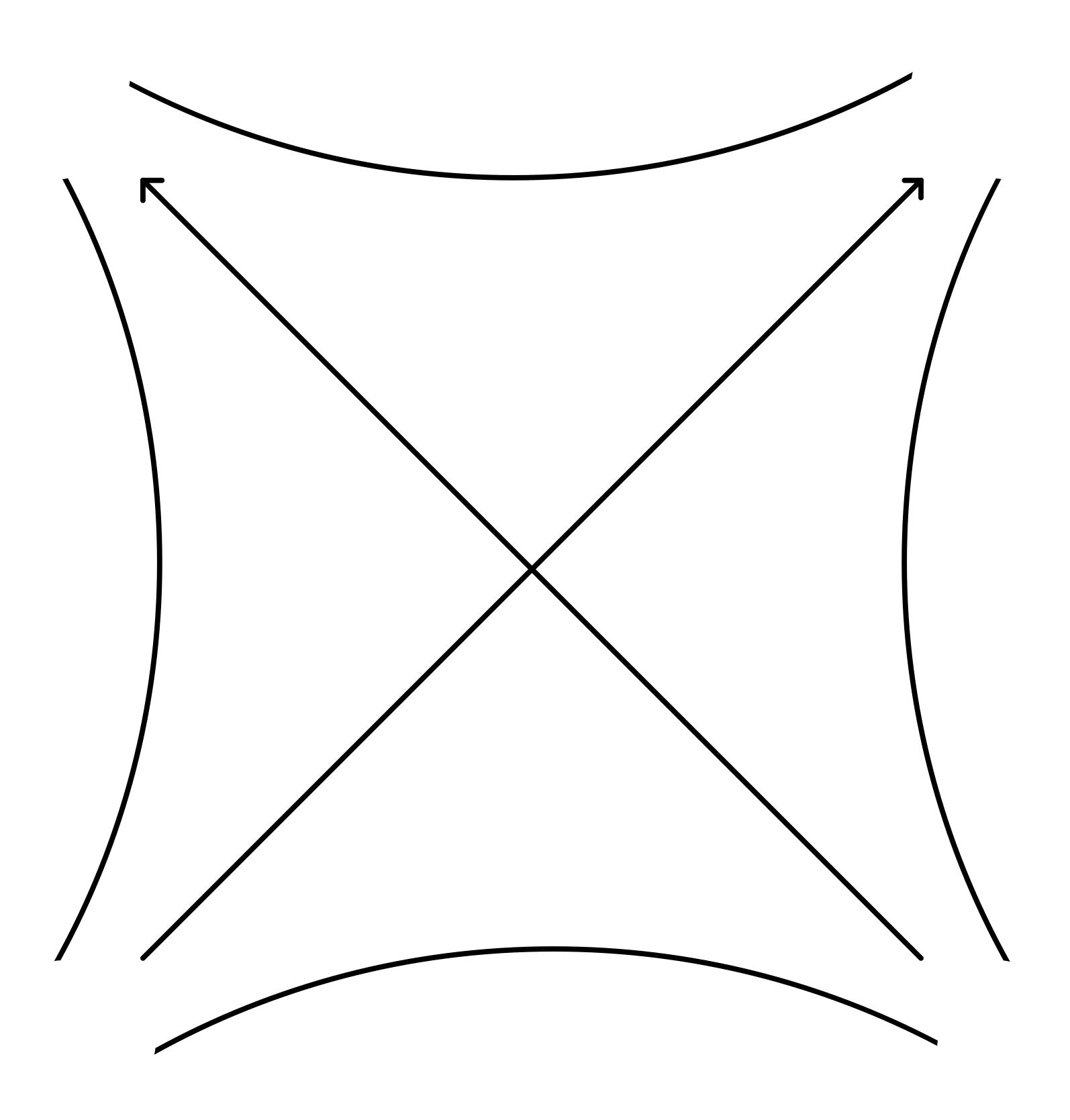}
			\put (84,83) {$x^{+}$}
			\put (8,83) {$x^{-}$}
			\put (93.5,55) {\footnotesize{$r=0$}}
			\put (91.5,50) {\footnotesize{$r_{*}=0$}}
			\put (85,45) {\footnotesize{$x^{+}x^{-}=-1/\Lambda$}}
			\put (85,40) {\footnotesize{`North pole'}}
			\put (-9,55) {\footnotesize{$r=0$}}
			\put (-11,50) {\footnotesize{$r_{*}=0$}}
			\put (-18,45) {\footnotesize{$x^{+}x^{-}=-1/\Lambda$}}
			\put (-19,40) {\footnotesize{`South pole'}}
			\put (49,47) {\rotatebox{45}{\footnotesize{Cosmological horizon}}}
			\put (45,52) {\rotatebox{45}{\footnotesize{$r=\ell$,\;$r_{*}=-\infty$,\;$x^{-}=0$}}}
			\put (19,95) {\footnotesize{$r=\infty$\,,\;\,$r_{*}=0\,,\;\,x^{+}x^{-}=1/\Lambda$}}
			\put (30,89) {\footnotesize{`Future infinity'}}
		\end{overpic}
	\end{center}
		\vspace{-0.25cm}
\caption{Conformal diagram for the three-dimensional de Sitter space, which is inherited by the two-dimensional de Sitter space. The transverse space at every point is a circle. The Kruskal coordinates in the north pole wedge are given in \eqref{eq:conf}.
}
\label{fig:conf1} 
\end{figure}

The three-dimensional action yielding pure de Sitter space is given by
\begin{equation}
	S_{\text{3d}}
	=
	\frac{1}{16\pi G_{\text{3d}}}\int d^{3}x\sqrt{-g_{\text{3d}}}\left[
		R_{\text{3d}}
		-
		2\Lambda
	\right]
	+
	\frac{1}{8\pi G_{\text{3d}}}\int d^{2}x\sqrt{-h_{\text{3d}}}K_{\text{3d}}
	\,,
\end{equation}	
where the Einstein Equations give us $R_{\text{3d}}=6\Lambda$ and the last term is the usual Gibbons-Hawking-York term. 
In Kruskal coordinates the three-dimensional de Sitter space metric is given by
\begin{equation}
	ds^{2}
	=
	\frac{1}{\left(1-\frac{ x^{+}x^{-}}{\ell^{2}}\right)^{2}}
	\left[
		-4dx^{+}dx^{-}
		+
		\left(1+\frac{ x^{+}x^{-}}{\ell^{2}}\right)^{2}\ell^{2}d\theta^{2}
	\right]
	\,,
\end{equation}
where $\theta$ has the range $[0,2\pi]$ and $-1/\Lambda \leq x^{+}x^{-}\leq 1/\Lambda$, see Figure \ref{fig:conf1} for a related conformal diagram. Plugging this metric into the Ricci scalar, we obtain $R_{\text{3d}}=6/\ell^{2}$ and we therefore conclude $\Lambda=\ell^{-2}$. Here $\ell$ represent the de Sitter length.
One can relate these Kruskal coordinates $x^{\pm}$ to the de Sitter space static patch coordinates $\sigma^{\pm}$, which cover the North pole wedge, via
\begin{equation}\label{eq:conf}
	x^{\pm}
	=
	\pm\frac{1}{\sqrt{\Lambda}}
	e^{\pm \sqrt{\Lambda} \sigma^{\pm}}
	\,,
	\quad
	\sigma^{\pm}
	=
	t\pm r_{*}
	\,,
\end{equation}
where $r_{*}$ is the tortoise coordinate in the static patch, which ranges from the cosmological horizon $r_{*}=-\infty$ (or in static patch coordinates $r=\ell$) to $r_{*}=0$, which is the location of the North pole (or in static patch coordinates $r=0$). Using Appendix \ref{sec:coordsandpatches} we can relate $r$ to $r_{*}$ and we can derive that the Gibbons-Hawking temperature is given by
\begin{equation}\label{eq:temp}
	T
	=
	\frac{\sqrt{\Lambda}}{2\pi}
	\,.
\end{equation} 
Using the Bekenstein-Hawking entropy one derives the entropy of de Sitter space in three dimensions to be \cite{PhysRevD.23.287,Gibbons:1977mu}
\begin{equation}\label{eq:entropy1}
	S
	=
	\frac{\pi }{2 G_{\text{3d}}\sqrt{\Lambda}}
	\,.
\end{equation}
For dimensional -- circular -- reduction we use the following Ansatz
\begin{equation}
	ds^{2}_{\text{3d}}
	=
	-e^{2\rho(x^{+},x^{-})}dx^{+}dx^{-}
	+
	\phi^{2}(x^{+},x^{-})d\theta^{2}
	\,,
\end{equation}
which will translate to Kruskal coordinates in two dimensions. For simplicity, we choose to put the Kaluza-Klein vector to zero.
This provides us with the following identities
\begin{equation}\label{eq:2dsols}
	\sqrt{-g_{\text{3d}}}
	=
	\sqrt{-g_{\text{2d}}}\phi
	\,,
	\quad
	\int d^{3}x
	=
	\int d^{2}x 2\pi
	\,,
	\quad
	R_{\text{3d}}
	=
	R_{\text{2d}}
	-
	\frac{2}{\phi}
	\square_{\text{2d}}\phi
	\,,
	\quad
	K_{3}
	=
	K_{2}
	+
	\frac{1}{\phi}\partial_{n}\phi
	\,,
\end{equation}
where $n$ denotes the normal.
Choosing $\Phi$ to be dimensionless, we introduce
\begin{equation}
	\frac{1}{4G_{\text{3d}}}\phi
	=
	\frac{1}{\pi}\Phi
	\,,
\end{equation}
which yields the dimensionally reduced action:
\begin{equation}\label{eq:theaction}
	S_{\text{2d}}
	=
	\frac{1}{2\pi}
	\int d^{2} x\sqrt{-g}\Phi\left[
		R
		-
		2\Lambda
	\right]
	+
	\frac{1}{\pi}\int dx\sqrt{-h}\Phi K 
	\,,
\end{equation}
where we dropped most subscripts referring explicitly to two dimensions and the `two-dimensional Newton's constant' can be read off to be $G_{\text{2d}}=1/8$. The bulk matches the Jackiw-Teitelboim (JT) model with an opposite sign in the potential \cite{JACKIW1985343,TEITELBOIM198341}.\footnote{If one were to start from a higher than three-dimensional Einstein-Hilbert action, one would get more terms in the final potential than only the JT term with the opposite sign. This is clear when considering the spherical reduction of a four-dimensional de Sitter space background, see e.g. \cite{Bousso:1997wi}, to arrive in two dimensions. If we then perform a Weyl rescaling in order to get rid of the kinetic term for the dilaton, the potential will have a term depending on $\Phi$ beyond the JT term. This would spoil the simple analytic behavior of JT.} By matching to the higher dimensional origin we find the following solutions to the fields
\begin{equation}
	ds^{2}_{\text{2d}}
	=
	-e^{2\rho}dx^{+}dx^{-}
	\,,
	\quad
	e^{2\rho}
	=
	\frac{4}{\left(1-\Lambda x^{+}x^{-}\right)^{2}}
	\,,
	\quad
	\Phi
	=
	\frac{S}{2}\frac{1+\Lambda x^{+}x^{-}}{1-\Lambda x^{+}x^{-}}
	\,,
\end{equation}
where three-dimensional entropy $S$ was introduced in \eqref{eq:entropy1}.
We check that the two-dimensional entropy coincides with the three-dimensional entropy by using the Bekenstein-Hawking formula (see e.g. \cite{Thorlacius:1994ip})
\begin{equation}\label{eq:bhf}
	\text{two-dimensional entropy}:=\left.\frac{\text{Area}}{4G_{\text{eff}}}
	\right|_{\text{horizon}}=2\Phi_{\text{horizon}}=S
	\,,
\end{equation} 
where the Area at the horizon is unity and $G_{\text{eff}}=G_{\text{2d}}/\Phi$.\footnote{Here $G_{\text{eff}}$ arises because the dilation mediates the gravitational coupling strength by appearing in front of the Ricci scalar in \eqref{eq:theaction}.} The temperatures in two and three dimensions also coincide. 


The here studied solution is different from the two-dimensional de Sitter space solutions studied in Refs. \cite{Chen:2020tes,Hartman:2020khs,Balasubramanian:2020xqf}, where islands were studied as well. Namely, the pure two-dimensional de Sitter space solution does not end at $x^{+}x^{-}=-1/\Lambda$, but rather extends further and even has a periodic identification, recall Figure \ref{fig:penroses} in the introduction. This causal structure is akin to the Schwarzschild-de Sitter black hole in e.g. four dimensions and can in fact be obtained by taking the black hole horizon close to the cosmological horizon followed by a near horizon limit. This gives one Nariai space see e.g. \cite{nariai1951new,Anninos:2012qw,Bousso:2002fq}, which, when spherically reduced, becomes the two-dimensional de Sitter space solutions studied, amongst others, in \cite{Chen:2020tes,Hartman:2020khs,Balasubramanian:2020xqf}. We stress, however, that in this work we focus on the causal structure of a higher dimensional pure de Sitter space instead, which we inherit from our explicit dimensional reduction that does not involve a black hole.

On a technical level this difference can be related to the fact that in e.g. Refs. \cite{Chen:2020tes,Hartman:2020khs,Balasubramanian:2020xqf} there is a topological term $\sim \int d^{2}x\sqrt{-g} S_{0}R$ added to the Lagrangian in \eqref{eq:theaction}. Here $S_{0}\gg S$ and its higher dimensional pedigree can often be interpreted as the entropy of an extremal limit, so in that case $\Phi$ controls the deviation from extremality. Although this term would not affect the equations of motion, it would in fact alter the effective Newton's constant and change the entropy formula in \eqref{eq:bhf} and rather than requiring $\Phi\geq0$, one has to require $S_{0}+\Phi\geq0$, which implies that $\Phi$ can be negative without any issue and can as such take larger domain which gives access to the full two-dimensional de Sitter. Requiring $S_{0}+\Phi\geq0$ translates to
\begin{equation}
	-\infty<x^{+}x^{-}<\infty\,,
\end{equation}
contrasting the case we consider in this paper where $\Phi\geq0$, which implies $-1/\Lambda\leq x^{+}x^{-}<\infty$. From here one can see that one can go `beyond' the pole at $-x^{+}x^{-}=1/\Lambda$ when a topological term of the type $\sim \int d^{2}x\sqrt{-g} S_{0}R$ is considered.
\subsection{Semi-classical corrections}
Let us consider the reduced model we obtained in the previous subsection. In the spirit of e.g. Refs. \cite{Callan:1992rs,Almheiri:2014cka,Christensen:1977jc}, we add conformal fields (CFT) with central charge $c$ and a semi-classical loop correction corresponding to the conformal anomaly known as the Polyakov term:
\begin{equation}
	S_{\text{2d}}
	=
	\frac{1}{2\pi}
	\int 
	d^{2}x \sqrt{-g}
	\left[
		\Phi 
		\left(
			R
			-
			2 \Lambda 
		\right)
		-
		\frac{c}{48}R\frac{1}{\nabla^{2}}R
	\right]
	+
	S_{\text{CFT}}
	\,,
\end{equation}
where we chose to drop the boundary terms without any loss of generality.
Due to the Polyakov term, there will be a non-zero energy flux that we will associate to radiation and it turns out that the solution for $\Phi$ will be back reacted.

The Polyakov term describes intrinsically two-dimensional radiation and is not the result from a reduction in three dimensions. One does not capture three-dimensional gray body factors, for instance. The addition of the Polyakov term can be viewed as a simplification of the problem at hand in favor of analytic control. Nevertheless, we stress that at leading the model is not modified if $S \gg c$. We furthermore require $c\gg1$, which is needed in order suppress any other loop order corrections. However, $c$ can not be so large that it dominates the classical term, which is proportional to entropy $S$. Large entropy $S$ can also be translated into $1/\sqrt{\Lambda}=\ell\gg G_{\text{3d}}$ which is the configuration in which classical contributions should indeed dominate.

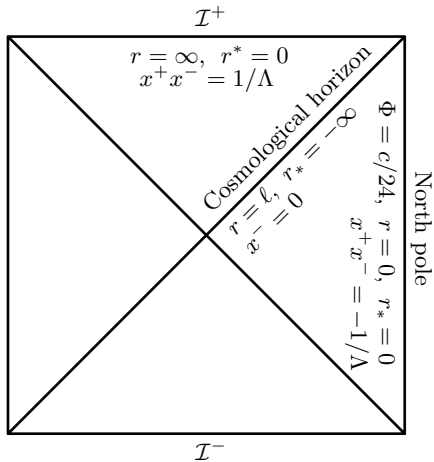
\begin{figure}[h!!]
	\begin{center}
		\begin{overpic}[width=0.4\textwidth]{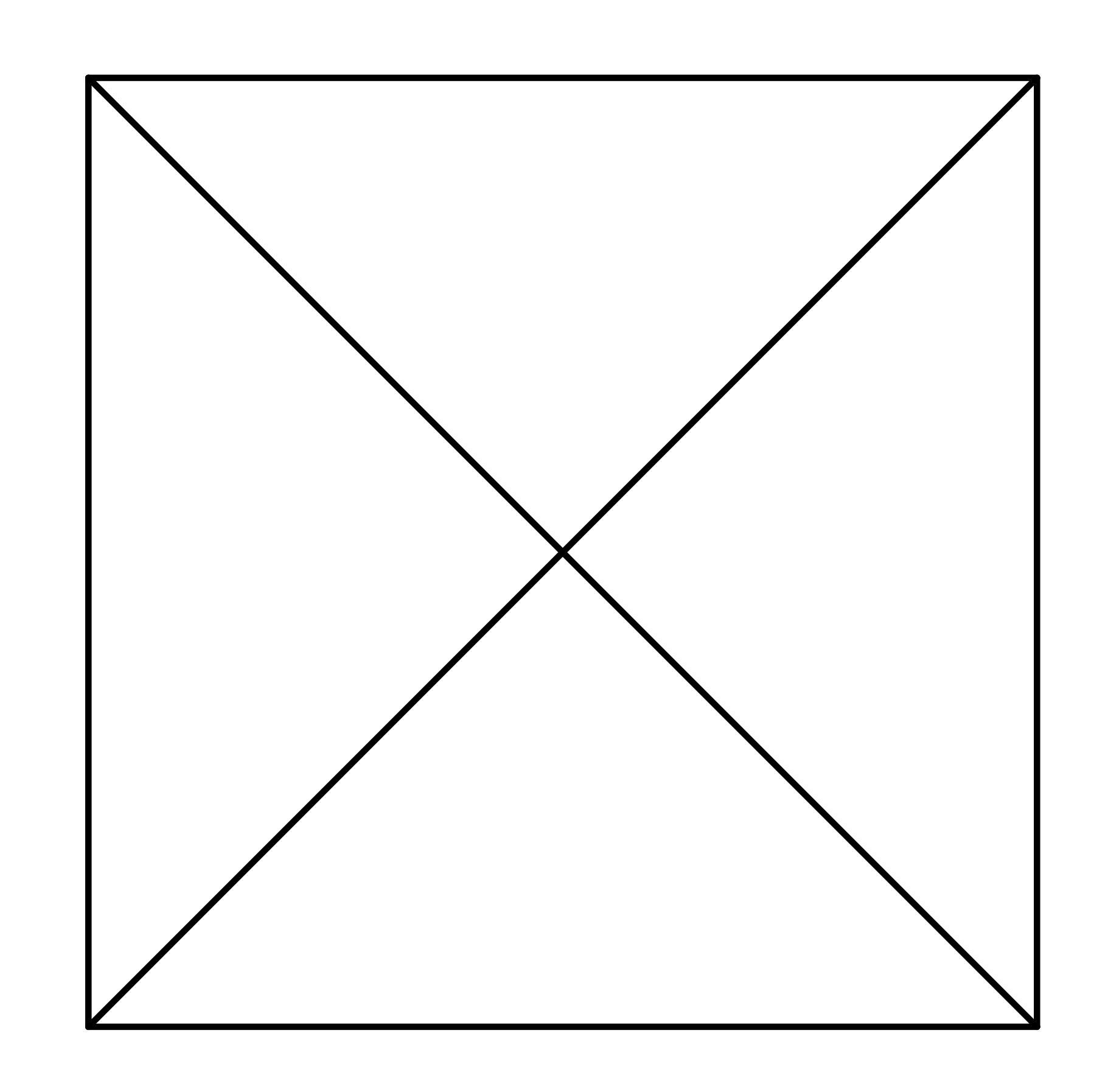}
			\put (34,85) {\footnotesize{$r=\infty$,\; $r^{*}=0$}}
			\put (36,80) {\footnotesize{$x^{+}x^{-}=1/\Lambda$}}
			\put (48,93) {\footnotesize{$\mathcal{I}^{+}$}}
			\put (48,0) {\footnotesize{$\mathcal{I}^{-}$}}
			
			\put (49,53) {\rotatebox{45}{\footnotesize{Cosmological horizon}}}
			\put (54,48) {\rotatebox{45}{\footnotesize{$r=\ell$,\; $r_{*}=-\infty$}}}
			\put (57,44) {\rotatebox{45}{\footnotesize{$x^{-}=0$}}}

			\put (86,77) {\rotatebox{-90}{\footnotesize{$\Phi=c/24$, \;$r=0$, \;$r_{*}=0$}}}
			\put (94,62) {\rotatebox{-90}{\footnotesize{North pole}}}
			\put (80,52) {\rotatebox{-90}{\footnotesize{$x^{+}x^{-}=-1/\Lambda$}}}
		\end{overpic}
	\end{center}
\caption{Penrose diagram of semi-classical de Sitter space with radiation in the North pole wedge defined with respect to the Hartle-Hawking vacuum. The dilaton $\Phi$ is weakly coupled in all these regions.
}
\label{fig:penrose1} 
\end{figure}

In Kruskal coordinates the equations of motion and constraints read:
\begin{equation}
	2\partial_{+}\partial_{-}\Phi
	-
	\Lambda \Phi e^{2\rho}
	=
	\pi T^{\text{CFT(classical)}}_{+-}
	+
	\pi T^{\text{Polyakov}}_{+-}
	\,,
\end{equation}
\begin{equation}
	-2e^{2\rho}
	\partial_{\pm}
	\left[
		e^{-2\rho}\partial_{\pm}\Phi
	\right]
	=
	\pi T^{\text{CFT(classical)}}_{\pm\pm}
	+
	\pi T^{\text{Polyakov}}_{\pm\pm}
	\,,
\end{equation}
\begin{equation}\label{eq:lagrange}
	R=2\Lambda
	\,,
\end{equation}
where the first and second equation come from the off-diagonal and diagonal part, respectively, of the Einstein equations. The third equation comes from the variation of $\Phi$. We assume that the equation of motion for the CFT is satisfied and that its energy-momentum tensor $T^{\text{CFT(classical)}}_{+-}=T^{\text{CFT(classical)}}_{\pm\pm}=0$.\footnote{A $c$ amount of free massless scalars would satisfy this requirement, see for instance \cite{Callan:1992rs}.}
It is important to stress that we can still excite the semi-classical part of the CFT, which will be reflected in the Polyakov term.
The energy-momentum tensor associated to the Polyakov term is given by
\begin{equation}\label{eq:semi-class}
	T^{\text{Polyakov}}_{\pm\pm}(x^{\pm})
	=
	\frac{1}{2\pi}
	\left\{
	\frac{c}{6}
	\left[
	\partial_{\pm}^{2}\rho
	-
	\partial_{\pm}\rho\partial_{\pm}\rho
	\right]
	-
	\frac{c}{12}
	t_{\pm}(x^{\pm})
	\right\}
	\,,
	\quad
	T^{\text{Polyakov}}_{+-}(x^{+},x^{-})
	=
	-
	\frac{1}{2\pi}\frac{c}{6}\partial_{+}\partial_{-}\rho
	\,,
\end{equation}	
where $t_{\pm}({x^{\pm}})$ will be explained below.
Note that if we take $c=0$, the semi-classical corrections disappear and we restore the classical result.
Using that the equation $R=2\Lambda$ in de semi-classical result is solved by $
	e^{2\rho}
	=
	\frac{4}{(1-\Lambda x^{+}x^{-})^{2}}
$ the relations in \eqref{eq:semi-class} simplify further to
\begin{equation}
	T^{\text{Polyakov}}_{\pm\pm}(x^{\pm})
	=
	-\frac{c}{24\pi}t_{\pm}(x^{\pm})
	\,,
	\quad
	T^{\text{Polyakov}}_{+-}(x^{+},x^{-})
	=
	-\frac{c}{12\pi}\frac{\Lambda}{(1-\Lambda x^{+}x^{-})^{2}}
	\,.
\end{equation}
We now consider $t_{\pm}(x^{\pm})$. These functions are interpreted as the flux measured by an observer at the North pole (see Figure \ref{fig:penrose1}) and depend on what we choose as our quantum mechanical vacuum in what coordinates, see e.g. \cite{Thorlacius:1994ip,hartman2015lectures}. Following \cite{Gibbons:1977mu}, we will make the choice that the observer experiences thermal equilibrium in the static patch, i.e. in the $\sigma^{\pm}$ coordinates. This translates into equal ingoing and outgoing flux at a temperature of $T=\sqrt{\Lambda}/(2\pi)$ (the Gibbons-Hawking temperature one expects from the metric, see Appendix \ref{sec:coordsandpatches}), which is also known as the Hartle-Hawking state.\footnote{In \cite{Aalsma:2019rpt} a setup was considered in which less radiation leaves the static patch than it enters. This is the Unruh state, which in our language would translate to $t_{+}(x^{+})\neq t_{-}(x^{-})$.} This defines the values for $t_{\pm}(\sigma^{\pm})$ in the following way. For a two-dimensional CFT on a plane, i.e. in the $\sigma^{\pm}$ coordinates, we know that $T^{\text{CFT}}_{\pm\pm}(\sigma^{\pm})=T^{2}\pi c/12$, see e.g. \cite{Datta:2019jeo}. Requiring the Gibbons-Hawking temperature $T=\sqrt{\Lambda}/(2\pi)$, we obtain $T^{\text{CFT}}_{\pm\pm}(\sigma^{\pm})=\frac{c}{24\pi }\frac{\Lambda}{2}$. This implies that $t_{\pm}(\sigma^{\pm})=-\Lambda/2$.

Starting from the relation between Kruskal coordinates $x^{\pm}$ and static patch coordinates $\sigma^{\pm}$ given in Eq. \eqref{eq:conf}, we can use the anomalous transformwation of the energy-momentum tensor via a Schwarzian derivative to find the following relation, where the accents indicate derivation with respect to $\sigma^{\pm}$,
\begin{equation}
	(x'^{\pm})^{2}t_{\pm}(x^{\pm})
	=
	t_{\pm}(\sigma^{\pm})
	-
	\left[
		\frac{x'''^{\pm}}{x'^{\pm}}
		-
		\frac{3}{2}
		\left(
			\frac{x''^{\pm}}{x'^{\pm}}
		\right)^{2}
	\right]
	\quad
	\Rightarrow
	\quad
	\Lambda(x^{\pm})^{2}t_{\pm}(x^{\pm})
	=
	t_{\pm}(\sigma^{\pm})
	+
	\frac{\Lambda}{2}
	\,,
\end{equation}
which means that $t_{\pm}(x^{\pm})=0$. This value in the $x^{\pm}$ coordinates is analogous to what happens in the semi-classical eternal black hole, see e.g. \cite{Almheiri:2019yqk,Callan:1992rs,Gautason:2020tmk,Hartman:2020swn}, which is also in thermal equilibrium. This result can also independently be reached using different approaches, see e.g. \cite{Aalsma:2019rpt}.

It turns out that only $\Phi$, which arises from solving Eq. \eqref{eq:lagrange}, receives a semi-classical correction
\begin{equation}
	\Phi
	=
	\frac{S}{2}\frac{1+\Lambda x^{+}x^{-}}{1-\Lambda x^{+}x^{-}}
	+
	\frac{c}{24}
	=
	\frac{S}{2}(1+\epsilon)
	\frac{1+\frac{1-\epsilon}{1+\epsilon}\Lambda x^{+}x^{-}}{1-\Lambda x^{+}x^{-}}
	\,,
\end{equation}
where we defined
\begin{equation}
    \epsilon
    :=
    \frac{c}{12S}
    \ll
    1
    \,.
\end{equation}
It remains true that the gravitational coupling is strong when $\Phi$ is near 0.\footnote{Following e.g. \cite{Susskind:1993if}, we can check that $\Phi=0$ still corresponds to the point where the gravitational coupling becomes strong, by computing the gravitational coupling constant directly from the action and by afterwards studying the prefactors of the kinetic terms in Kruskal coordinates. While for simplicity introducing $\Phi=e^{-\phi}$, we can then group the kinetic terms in the action as $\partial_{+}\chi \cdot M \cdot \partial_{-}\chi$ where $\chi =(\phi,\rho)$. We find
\begin{equation}
	M
	=
	\begin{pmatrix}
		0&2\Phi\\
		2\Phi&-\frac{c}{6}
	\end{pmatrix}
	\,.
\end{equation}
The following expression can be interpreted as the gravitation coupling constant: $[-(\det M)/4]^{-1/4}=\Phi^{-1/2}$, which indeed confirms that gravitational quantum corrections become strong near $\Phi=0$.} 
However, now $\Phi$ goes to zero when $x^{+}x^{-}=-\frac{1+\epsilon}{1-\epsilon}1/\Lambda$, which is outside the range of what the coordinates cover and we thus do not have to worry about coupling becoming strong in regions of our interest. The resulting semi-classical Penrose diagram is presented in Figure \ref{fig:penrose1}.

\section{Evaluating the generalized entropy}\label{sec:pagecurve}
Our anchor curve will be hugging the cosmological horizon, as discussed in Section \ref{sec:motivation}, which will be made more precise in this Section. This allows us to consider the generalized entropy between the states in the interior of the cosmological horizon and the complement, see Figure \ref{fig:island1} below. It will turn out that the generalized entropy grows initially, but saturates at a value of the generalized entropy corresponding to the value of the Gibbons-Hawking entropy, see Figure \ref{fig:pagecurve}.
\subsection{The generalized entropy}
The formula for the generalized entropy is given by \cite{Ryu:2006bv,Hubeny:2007xt,Faulkner:2013ana,Engelhardt:2014gca,Penington:2019npb,Almheiri:2019hni}\begin{equation}\label{eq:genent1}
	S_{\text{gen}}
	=
	\frac{\text{Area}(I)}{4G_{\text{eff}}}
	+
	S_{\text{Bulk}}[\mathcal{S}_{AI}]
	\,,
\end{equation}
where $I$ is the location of the extremal surface (just a point in our case), which is found by extremizing $S_{\text{gen}}$, and $S_{\text{Bulk}}[\mathcal{S}_{AI}]$ is the von Neumann entropy of the radiation in spacelike surface $\mathcal{S}_{AI}$, which lies between $I$ and anchor point $A$, see Figure \ref{fig:island1}. Using the generalized entropy one is instructed to compare the island and the no island ($I$ extending all the way to the South pole) configurations and pick the one with the lowest generalized entropy at each instance of time. This competition is the mechanism which will give us a transition in the behavior of the entanglement entropy growth. We will now assemble \eqref{eq:genent1} into an explicitly usable form in the current setting.

In equation \eqref{eq:bhf} we concluded that
\begin{equation}
	\frac{\text{Area}(I)}{4G_{\text{eff}}}
	=
	2\Phi(I)
	\,.
\end{equation}
In order to compute $S_{\text{Bulk}}[\mathcal{S}_{AI}]$ we adopt the approach of \cite{Almheiri:2019hni} where the leading order Ryu-Takayanagi formula is applied \cite{Ryu:2006bv}. In other words, we will relate $S_{\text{Bulk}}[\mathcal{S}_{AI}]$ to a geodesic between points $A$ and $I$ once the CFT is embedded in $\text{AdS}_{3}$.
In order to embed our problem into $\text{AdS}_{3}$ we consider $S_{\text{Bulk}}[\mathcal{S}_{AI}]$ on flat space instead of curved spacetime by performing a Weyl rescaling with which we can get rid of the conformal factor $e^{2\rho}$ in front of the metric (3.7.). 
We obtain
\begin{equation}
	ds^{2}_{\text{3d}}
	=
	\frac{\ell_{\text{3d}}^{2}}{z^{2}}
	\left(
		dz^{2}
		-
		dx^{+}dx^{-}
	\right)
	\,,
\end{equation}
where $z$ is the holographic direction, $\ell_{\text{3d}}$ is the intrinsic length scale of $\text{AdS}_{3}$ and we use the CFT vacuum coordinates $x^{\pm}$. The well-known geodesics corresponding to this problem are half circles which pierce into the interior of $\text{AdS}_{3}$ connecting the boundary points $(x^{+}_{A},x^{+}_{A},z_{A})$ 
and $(x^{+}_{I},x^{+}_{I},z_{I})$. If we just consider $z_{I}=\delta_{I}$ and $z_{A}=\delta_{A}$, where the $\delta$s represent some cut-off length scale, we would arrive at the usual answer (where we use Brown-Henneaux to relate $\ell_{\text{3d}}$ and the three dimensional Newton's constant to the central charge on the boundary \cite{Brown:1986nw})
\begin{equation}
	\left.
	S_{\text{Bulk}}[\mathcal{S}_{AI}]\right|_{\text{flat space}}
	=
	\frac{c}{6}\log
	\left[
		\frac{d(A,I)^{2}}{\delta_{I}\delta_{A}}
	\right]
	\,.
\end{equation} 
Here $d(A,I)^{2}=(x^{+}_{A}-x^{+}_{I})(x^{-}_{A}-x^{-}_{I})$ is the squared distance between the points $A$ and $I$ expressed in $ds^{2}=-dx^{+}dx^{-}$. In order to obtain the curved space result we implement a coordinate transformation on the holographic boundary through $z_{A}=\delta e^{-\rho_{A}}$ and $z_{I}=\delta e^{-\rho_{I}}$, which undoes the earlier applied Weyl transformation. Here $\rho_{A}$ and $\rho_{I}$ are the conformal factor evaluated respectively at the point $A$ and $I$. The final result becomes
\begin{equation}
	S_{\text{Bulk}}[\mathcal{S}_{AI}]
	=
	\frac{c}{12}\log
	\left[
		(x^{+}_{A}-x^{+}_{I})^{2}(x^{-}_{A}-x^{-}_{I})^{2}e^{2\rho_{A}}e^{2\rho_{I}}
	\right]
	\,.
\end{equation} 
It is important to note that throughout this computation the gravitational coupling related to $S_{\text{Bulk}}[\mathcal{S}_{AI}]$ is small since we assumed $S\gg c\gg1$ and as a result we do not expect complications with this Planck brane and its extension into $\text{AdS}_{3}$.
More computational details on the here followed approach can be found in e.g. \cite{Gautason:2020tmk}. 
This same result can also found using methods described in Ref. \cite{Fiola:1994ir} without using the AdS/CFT duality.
Plugging the expressions for the von Neumann entropy and the area term into \eqref{eq:genent1} and adopting the rescaled coordinates $v=\sqrt{\Lambda}x^{+}$ and $u=\sqrt{\Lambda}x^{-}$, we explicitly obtain
\begin{equation}\label{eq:finalsgen}
	S_{\text{gen}}/S
	=
	(1+\epsilon)
	\frac{1+\frac{1-\epsilon}{1+\epsilon} v_{I} u_{I}}{1- v_{I} u_{I}}
	+
	\epsilon
	\log
	\left[
		16\frac{
			(v_{A}-v_{I})^{2}(u_{A}-u_{I})^{2}
		}
		{
			(1-v_{I} u_{I})^{2}(1-v_{A} u_{A})^{2}
		}
	\right]
	\,.
\end{equation}
Typically we consider the time coordinate of an anchor point on the anchor curve, $t_{A}$, at late time, i.e. $\sqrt{\Lambda}t_{A}\sim 1/\epsilon$, where $\epsilon=c/(12S)\ll 1$. The spatial location of the anchor curve, in terms of the tortoise coordinate $r_{*}$, we choose $r_{A}\sim \frac{1}{\sqrt{\Lambda}}\log \delta$. We take $\delta\ll\epsilon$ such that the anchor curve will be hugging the cosmological horizon. This $\delta$ can be interpreted as analogous to the location of the anchor curve $\Lambda_{\text{anchor}}$ in the flat space setup \cite{Gautason:2020tmk}, for which one required $1/\Lambda_{\text{anchor}}\ll \epsilon$.
\subsection{Island, no island, and the Page curve}
\subsubsection*{Island scenario} 
Let us consider the case in which there is an island. In order to find the location of the island, we have to extremize the generalized entropy \eqref{eq:finalsgen} with respect to $v_{I}$ and $u_{I}$, which respectively gives us
\begin{equation}
    \frac{
        u_{I}(v_{A}-v_{I})
        +
        \epsilon
        \left(
	    (v_{A}-v_{I})(1-u_{I}v_{I})u_{I}
            -
            (1-v_{I}u_{I})^{2}
        \right)
}{(1-u_{I}v_{I})^{2}(v_{A}-v_{I})}
    =
    0
    \,,
\end{equation}
\begin{equation}
 \frac{
        v_{I}(u_{A}-u_{I})
        +
        \epsilon
        \left(
            (u_{A}-u_{I})(1-v_{I}u_{I})v_{I}
            -
            (1-v_{I}u_{I})^{2}
        \right)
   }{(1-u_{I}v_{I})^{2}(u_{A}-u_{I})}
    =
    0
    \,.
\end{equation}
This yields three lengthy analytic solutions. One of these solutions lies beyond the allowed range of $x^{+}x^{-}\geq -1/\Lambda$. Another solution lies to the right hand side of the anchor curve and is therefore discarded. At leading order, we are left the solution
\begin{equation}
	v_{I}
	\approx
	-\sqrt{\epsilon}e^{\sqrt{\Lambda}t_{A}}
	\,,
	\quad
	u_{I}
	\approx
	\sqrt{\epsilon}e^{-\sqrt{\Lambda}t_{A}}
	\,,
\end{equation}
where we assumed $r_{A}=\frac{a}{\sqrt{\Lambda}} \log \epsilon$, with $a$ some positive real number that drops out of the result at this order.
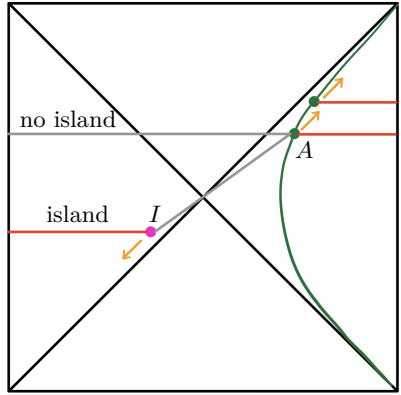
\begin{figure}[h!!]
	\begin{center}
		\begin{overpic}[width=0.38\textwidth
		]{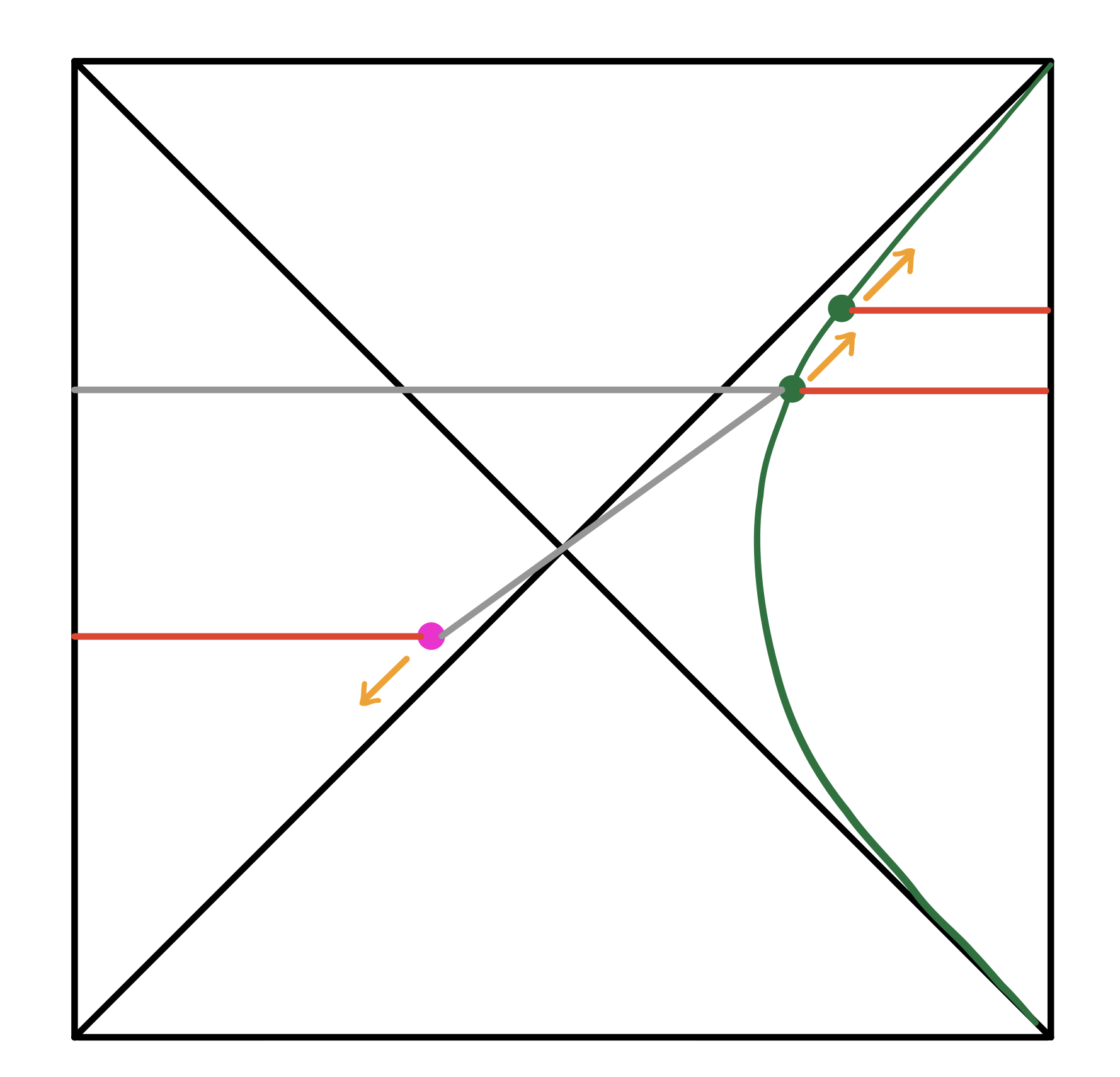}
			\put (15,41.5) {\footnotesize{island}}
			\put (38,41.5) {\footnotesize{$I$}}
			\put (9,63) {\footnotesize{no island}}
			\put (71,56) {\footnotesize{$A$}}
		\end{overpic}
	\end{center}
	\vspace{-0.25cm}
\caption{
The green curve represents the anchor curve which divides the system into the interior of the cosmological horizon (red line) and its complement. The green dots are anchor points ($A$) and the pink dot the extremal surface ($I$). After some time there is a transition of dominance of the `no island' solution and the `island' solution.
}
\label{fig:island1} 
\end{figure}

We present the island scenario in Figure \ref{fig:island1}. Notice that the island runs downwards, as time $t_{A}$ on the anchor curve increases. Filling out the location of the extremal surface into the generalized entropy formula \eqref{eq:finalsgen} gives us
\begin{equation}
    S_{\text{island}}/S
    =
    1
    +
    \mathcal{O}(\epsilon)
    \,.
\end{equation}

\subsubsection*{No island scenario} 
In the case that there is no island, we have to study the trivial surface $\emptyset$. This means that $I$ is located on the South pole. This requires us to put $v_{I}=v_{0}$ and $u_{I}=u_{0}$, where $v_{0}<0$ and $u_{0}>0$ are fixed points for which $v_{0}u_{0}=-1$. The generalized entropy is not sensitive to further details of $v_{0}$ and $u_{0}$ at leading order in $\epsilon$:
\begin{equation}
	S_{\text{no island}}/S
	=
    2\epsilon t_{A} \sqrt{\Lambda}+
	\mathcal{O(\epsilon)}
    \,.
\end{equation}

\subsubsection*{Page curve} 
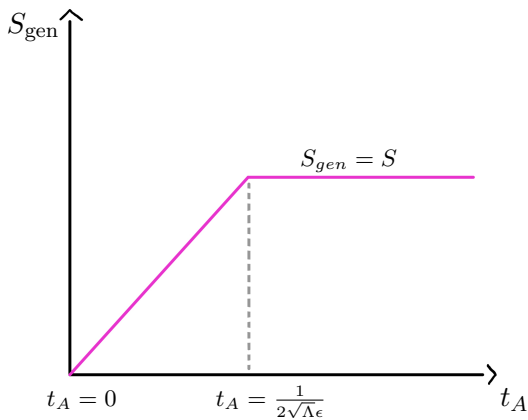
\begin{figure}[h!!]
	\begin{center}
		\begin{overpic}[width=0.38\textwidth, 
		]{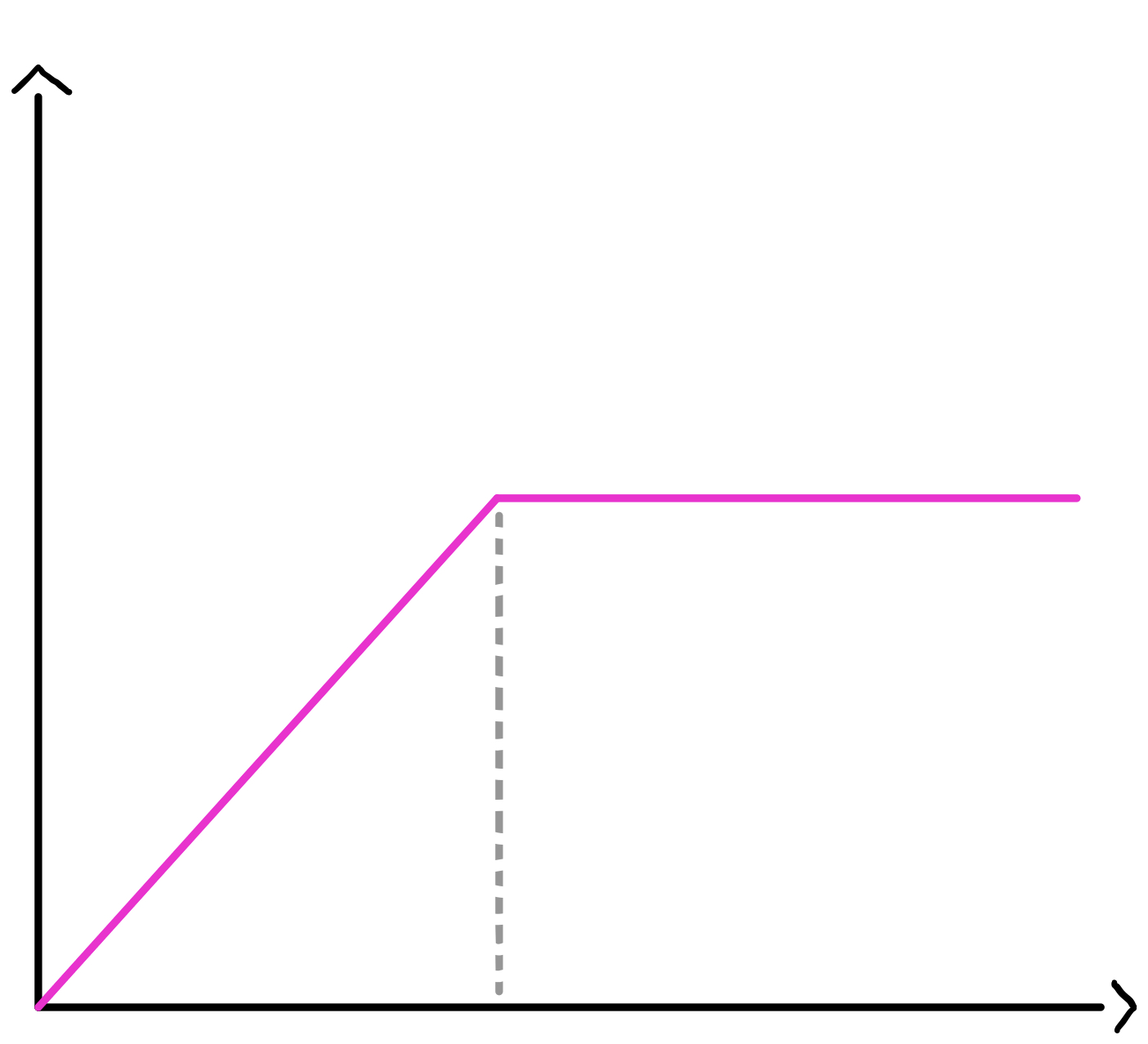}
			\put (-2,-4) {\footnotesize{$t_{A}=0$}}
			\put (36,-4) {\footnotesize{$t_{A}=\frac{1}{2\sqrt{\Lambda}  \epsilon}$}}
			\put (100,-4) {$t_{A}$}
			\put (-11,80) {$S_{\text{gen}}$}
			\put (55,50) {\footnotesize{$S_{gen}=S$}}
		\end{overpic}
	\end{center}
\caption{
The Page curve for a static patch in de Sitter space.
}
\label{fig:pagecurve} 
\end{figure}

We can now evaluate $S_{\text{gen}}=\text{min}(S_{\text{island}},S_{\text{no island}})$. We plot this in Figure \ref{fig:pagecurve}. At Page time $t_{A}=1/(2 \sqrt{\Lambda} \epsilon)$ there is a transition from the growing behavior to capping off at $S_{\text{gen}}=S$. 

In this setup, $t_{A}=0$ is some arbitrary moment. One could also consider a setup in which one e.g. tunnels into this state \cite{PhysRevD.28.2960} or forms de Sitter space through gravitational collapse, in which case $t_{A}=0$ would be a meaningful moment. Such models are not considered here and as such the important message is that the $S_{\text{gen}}$ is actually bounded from above by the Gibbons-Hawking entropy $S$. Furthermore, as $\Lambda\to0$, which is the limit in which we return to flat space and the size of the cosmological horizon blows up, indeed, the page curve predicts that entanglement entropy between the interior of the cosmological horizon and `what lies beyond' will grow forever. 
\section{The island moving back in time and entanglement wedge reconstruction}\label{sec:discussion}
\subsection{Island location}
The island in de Sitter space is far away -- unreachable in fact -- from the future cosmological horizon in which somebody from the North pole could throw something. This is in contrast to black holes in e.g. flat space and Anti-de Sitter space, see e.g. \cite{Penington:2019npb,Almheiri:2019psf,Gautason:2020tmk,Almheiri:2019yqk}, where the island is always found close to this future horizon. This difference in location is important, because otherwise you could harm the no-cloning theorem when combined with entanglement wedge reconstruction, as we will argue below.

\begin{figure}[h!!]
	\begin{center}
		\begin{overpic}[width=0.90\textwidth]{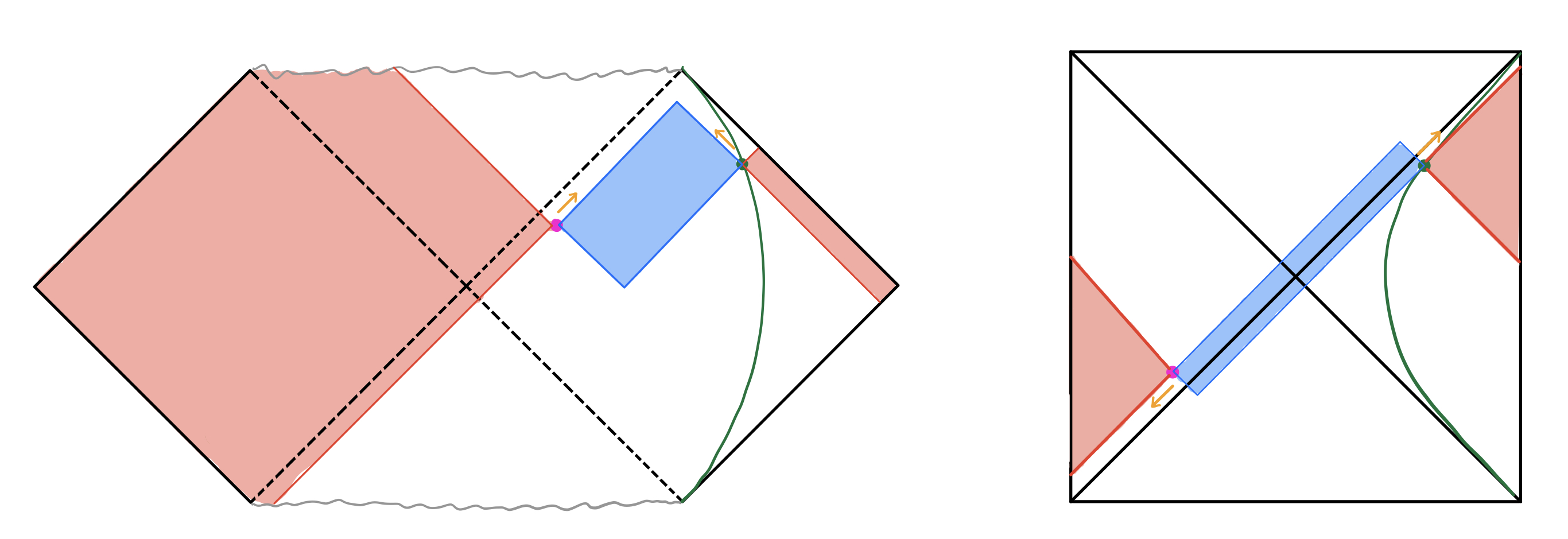}
		\end{overpic}
	\end{center}
	\vspace{-0.25cm}
\caption{In both figures the pink dot indicates the location of the extremal surface, the green dot indicates the location of the anchor point on the anchor curve (green line). The orange arrows denote movement directions in time. Left: A Penrose diagram of an eternal black hole in flat space is shown \cite{Gautason:2020tmk} where the gray wavy line denotes the location of the black hole singularity. As soon as a diary enters the red entanglement wedge within the event horizon (dashed lines), someone on the right hand side of the anchor curve, in the red entanglement wedge, can in principle reconstruct this diary. For generic black holes the extremal surface will be just inside or outside the future event horizon, allowing for reconstruction of things falling in. Right: Penrose diagram of pure de Sitter space beyond two dimensions. Due to the location of the entanglement wedges, someone within the North pole wedge will be unable to reconstruct things falling into the future cosmological horizon of the North pole wedge.
}
\label{fig:bhvsds1} 
\end{figure}

The idea of entanglement wedge reconstruction suggests, see e.g. \cite{Penington:2019npb}, that having an island near the future horizon implies the possibility of using Hawking radiation for decoding something that falls through the horizon. This is true because the entanglement wedge of the island is associated with the entanglement wedge on the right hand side of the anchor curve, i.e. the region of all the states in some observer's cosmological bubble at some moment, see Figure \ref{fig:bhvsds1} for more details. 
In the case of the black hole, if something falls into the horizon, at some moment it will be in the entanglement wedge of the island and therefore reconstructable for an observer who collects radiation in the corresponding entanglement wedge outside the horizon. The island in de Sitter space tells us that one will not be able to reconstruct things that fall into the cosmological horizon.

We can verify this island behavior in de Sitter space by doing a Gedanken experiment which does not involve islands. 
If one were able to reconstruct the diary from radiation in de Sitter space, the following could happen. Bob, who sits on the North pole, tosses his diary through his cosmological horizon. Alice, who sits in Bob's North pole wedge, starts to collect radiation and at some moment, after some time called scrambling time, she has reconstructed Bob's diary entirely. She jumps into the future cosmological horizon of Bob and catches up with the diary and thus possesses two exact versions of this diary, which violates the no-cloning theorem. In the case of the black hole, it can be shown that if Alice waits for scrambling time, the diary will inevitably fall into the singularity before Alice can chase it down, see e.g. \cite{Hayden:2007cs,Sekino:2008he}. The no-cloning theorem is preserved.

In other words: in absence of a singularity in de Sitter space, the island has to be where it is, because otherwise Alice can reconstruct the diary and then has all the time in the world to chase after the original diary behind the horizon in order to violate the no-cloning theorem.
\subsection{Island moving back in time}
Another striking difference with islands arising in black holes, is that the island in de Sitter space travels back in time, as $t_{A}$ increases. It is important that the island runs back in time, as will be clear from the following Gedanken experiment. 
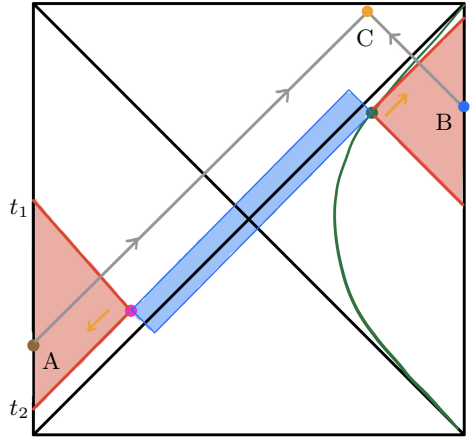
\begin{figure}[h!!]
	\begin{center}
		\begin{overpic}[width=0.40\textwidth]{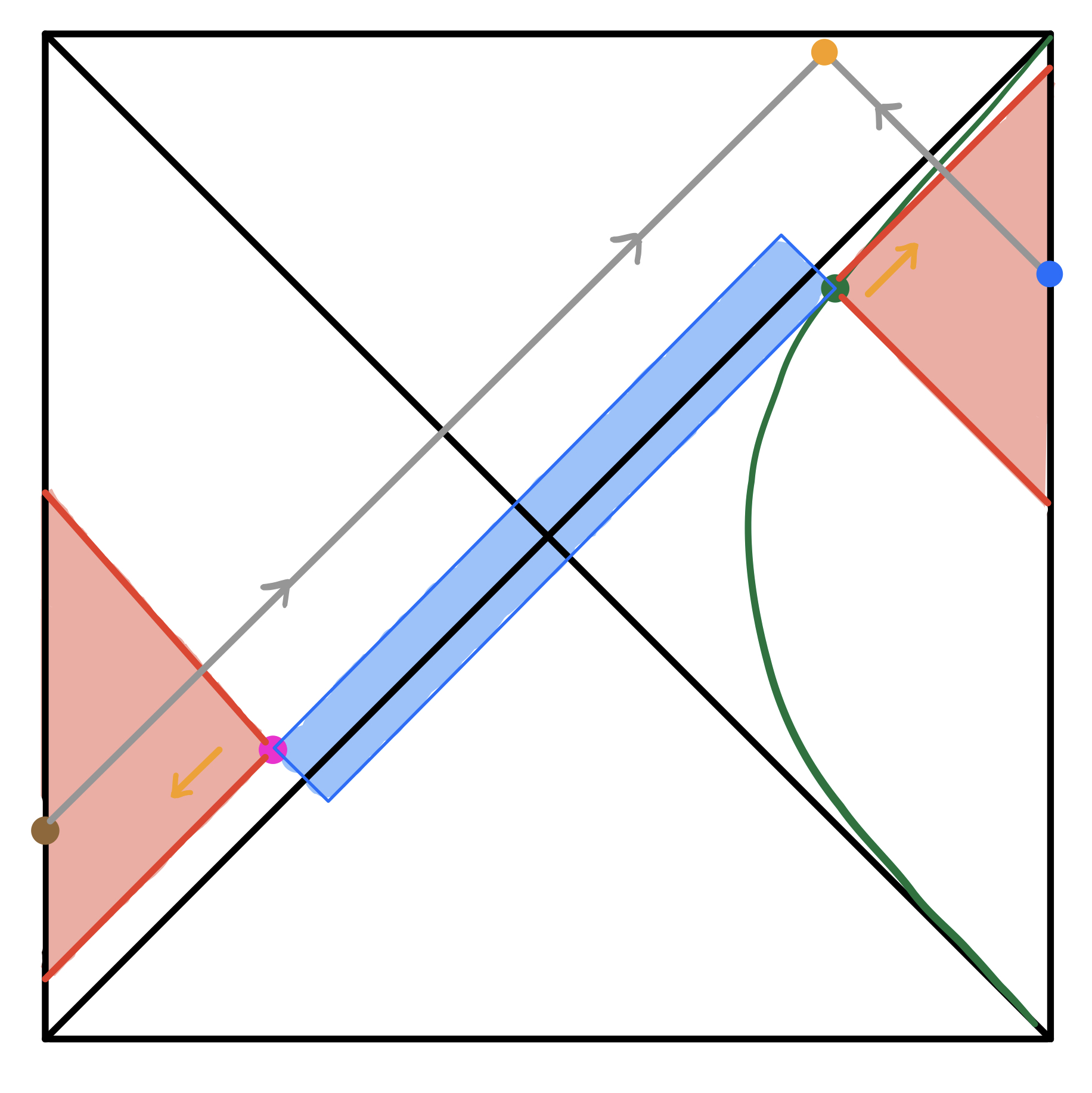}
			\put (73,88) {\footnotesize{C}}
			\put (90,70) {\footnotesize{B}}
			\put (6,19) {\footnotesize{A}}
			\put (-1,52) {\footnotesize{$t_{1}$}}			
			\put (-1,9) {\footnotesize{$t_{2}$}}		
		\end{overpic}
	\end{center}
	\vspace{-0.25cm}
\caption{A Penrose diagram of de Sitter space is shown. The green line is the anchor curve and the green point is the anchor point. The pink dot represent the extremal surface and the red and blue wedges are entanglement wedges. The red entanglement wedge of the island and the red entanglement wedge on the right hand side of the anchor curve are associated with each other as discussed in e.g. \cite{Penington:2019npb}.
}
\label{fig:scramble1} 
\end{figure}

Let us again place Bob (B) with his diary on the North pole. We put Alice (A) on the south pole in the entanglement wedge of the island, see Figure \ref{fig:scramble1} for the setup. 
As long as Alice and Bob are in the entanglement wedge, we assume that Alice is able to reconstruct Bob's diary (who has not tossed it through his horizon yet) from collecting radiation. As soon as either of them move out of the wedge, it is impossible to reconstruct.  

If Alice succeeds in reconstructing Bob's diary there would be a scenario possible in which they both toss their diaries (i.e. Bob his own diary and Alice her copy of Bob's diary) into their own cosmological horizon and Charlie, who lives beyond both cosmological horizons, can then compare both diaries and conclude that the no-cloning theorem has been violated. To make sure this does not happen, we have to limit the time Alice can spend within the entanglement wedge and as a result it has to be that the island moves back in time, instead of moving in the same direction as Alice (which is upwards in the Penrose diagram). However, Alice can still spend a finite amount of time within the wedge. It therefore has to be, in order to preserve the no-cloning theorem, that within this timescale decoding should be impossible. 

We can compute this timescale. Using $t=\frac{1}{2\sqrt{\Lambda}} \log (-u/v)$ we can compute the time difference between $t_{1}$ and $t_{2}$ (see Figure \ref{fig:scramble1}). In analogy to the Hayden-Preskill protocol for black holes \cite{Hayden:2007cs}, this will encode the scrambling time $t_{s}$, the minimal time it takes to be able to reconstruct something in such a setup, in order to protect the no-cloning theorem. It is found that for pure de Sitter, the scrambling time is
\begin{equation}
	t_{s}
	\sim
	\frac{1}{2\pi T}\log S
	+
	\mathcal{O}(\epsilon)
	\,,
\end{equation}
where we inserted temperature $T$ in favor of $\Lambda$ using \eqref{eq:temp}.
This coincides with the scrambling time value computed for black holes. If we would have taken $r_{A}$ to go to $-\infty$ slower than $\sim\log\epsilon$, e.g. $r_{A}\sim - c$, then $t_{s}\sim - c$. However, the point is that $\sim\frac{1}{2\pi T}\log S$ is the maximal time scale for which one minimally has to wait before being able to reconstruct something from the radiation.

These Gedanken experiments do not rely on specific dimensions. We speculate that the same qualitative island behavior should hold for any pure higher dimensional de Sitter space as well.

\section{Outlook}\label{sec:outlook}
In this paper we used two-dimensional methods to study islands in a model that has pure three-dimensional de Sitter space as its solution. We used the island formula to find that the entanglement entropy of the static patch cannot grow beyond the Gibbons-Hawking entropy. This island is furthermore found to behave differently from islands of black hole horizons in e.g. Anti-de Sitter space and flat space because 1) it moves back in time and 2) it sits in a different quadrant of the Penrose diagram. 
We verify that this has to be the case, because otherwise the no-clone theorem would be harmed in Hayden-Preskill like setups \cite{Hayden:2007cs,Susskind:1993mu} when combined with entanglement wedge reconstruction.

It seems reasonable, especially aided by the Gedanken experiments, to speculate a same type of qualitative island behavior beyond three dimensions. We furthermore expect the scrambling time to generalize to higher dimensions as well. 

It should be emphasized that the usage of the island description in this paper is speculative, although the results it gives seem reasonable.  
The approach employed in the current paper does not make any specific use of the Penrose diagram quadrant that contains $\mathcal{I}^{+}$, which is the location of the conjectured dual of de Sitter space, see e.g. \cite{Strominger:2001pn}. 
We argue however that in the current setup we are not interested in `storing' radiation and have reasonable weak gravity in the areas of interest.
From that perspective it would be interesting to try to motivate the here performed analysis in some holographic setting in the hope that it might prove insights about de Sitter holography. A worthwhile starting point could be the $\text{dS/dS}$ correspondence \cite{Alishahiha:2004md}. Beyond holography a Euclicean approach involving wormhole saddle points could be an insightful way of justifying the usage of the island formula in the here describe manner.

Apart from entropy and temperature, we see that the cosmological horizon shares features with black holes such as a Page curve and scrambling time. What about the status of holographic complexity of pure de Sitter space? Employing the here used approach of anchor curves, one can revisit de Sitter complexity, see e.g. \cite{Reynolds:2017lwq}, along the lines of \cite{Schneiderbauer:2019anh,Schneiderbauer:2020isp}, for potential new insights.
\\
\\
\textbf{Acknowledgments.} The author thanks Fri\dh rik Freyr Gautason and L\'{a}rus Thorlacius for their extensive contributions to this project. It is furthermore a pleasure to thank Lars Aalsma, Nick Poovuttikul, Lukas Schneiderbauer, and Manus Visser for stimulating discussions and comments on the draft. This research is supported by the Icelandic Research Fund (IRF) via a Personal Postdoctoral Fellowship Grant (185371-051).
\appendix
\section{de Sitter space in different coordinates}\label{sec:coordsandpatches}
In this Appendix we establish the connection between the static patch radial coordinate $r$ and the tortoise coordinate $r_{*}$. We furthermore explicitly go from the static patch coordinates to the Kruskal coordinates.

In the static patch we write the coordinates as 
\begin{equation}
	ds^{2}
	=
	-f(r)dt^{2}
	+
	\frac{dr^{2}}{f(r)}
	=
	f(r)
	\left[
		-dt^{2}
		+
		dr_{*}^{2}
	\right]
	\,.
\end{equation}
Solving the Einstein equations we find, where $\ell$ is the de Sitter length and $\Lambda=1/\ell^{2}$,
\begin{equation}
	f(r)
	=
	c_{1}
	+
	c_{2}r
	-\frac{ r^{2}}{\ell^{2}}
	\,.
\end{equation}
We will not specify $c_{1}$ and $c_{2}$ to remain fully general.
Here the radial coordinate $r$ has horizons at
\begin{equation}
	r_{\pm}
	=
	\frac{c_{2}}{2}\ell^{2}
	\pm
	\ell
	\sqrt{
		c_{1}
		+
		\ell^{2}
		\left(
			\frac{c_{2}}{2}
		\right)^{2}
	}
	\,.
\end{equation}
For the tortoise coordinate we find
\begin{equation}
	r_{*}
	=
	\frac{\ell}{\sqrt{\frac{c_{1}}{\ell}+\ell\left(\frac{c_{2}}{2}\right)^{2}}}
	\text{ArcTanh}
	\left[
		\frac{r-\ell^{2}\frac{c_{2}}{2}}{\sqrt{c_{1}+\ell^{2}\left(\frac{c_{2}}{2}\right)^{2}}}
	\right]
	\,,
\end{equation}
which is invertible with respect to $r$.
If we define 
\begin{equation}
	V
	=
	e^{\frac{\sqrt{c_{1}+\left(\frac{c_{2}\ell}{2}\right)^{2}}}{\ell}(t+r_{*})}
	\,,
	\quad
	U
	=
	-
	e^{\frac{\sqrt{c_{1}+\left(\frac{c_{2}\ell}{2}\right)^{2}}}{\ell}(-t+r_{*})}
	\,,
\end{equation}
we recover Kruskal coordinates
\begin{equation}
	ds^{2}
	=
	-
	\frac{4\ell^{2}}{(1-UV)^{2}}dUdV
	\,.
\end{equation}

\bibliographystyle{JHEP}
\bibliography{refds}
\end{document}